\documentclass[proceedings, preprint]{rmaa_2}

\usepackage{paralist}

\usepackage{psfrag,color}

\SetYear{2009}
\SetConfTitle{The Interferometric View on Hot Stars}

\title{Interferometric Studies of Hot Stars at Sydney University} 

\author{
  J. G. Robertson,\altaffilmark{1} 
  J. Davis,\altaffilmark{1}
  M. J. Ireland,\altaffilmark{1}
  P. G. Tuthill,\altaffilmark{1}
  W. J. Tango,\altaffilmark{1}
  A. P. Jacob,\altaffilmark{1}
  J. R. North\altaffilmark{1}
  and T. A. ten Brummelaar\altaffilmark{2}}

\altaffiltext{1}{Sydney Institute for Astronomy, School of Physics,
  University of Sydney, NSW 2006 Australia
  (g.robertson@physics.usyd.edu.au).} 

\altaffiltext{2}{Center for High Angular Resolution Astronomy, Georgia
  State University, P.O. Box 3969, Atlanta, GA 30302-3969
  USA(theo@chara-array.org).} 

\shortauthor{Robertson et al.}
\shorttitle{Interferometry at Sydney University}

\listofauthors{ J. G. Robertson, J. Davis, M. J. Ireland, P. G. Tuthill, 
W. J. Tango, A. P.Jacob, J. R. North, T. A. ten Brummelaar}
\indexauthor{Robertson, J.G.}
\indexauthor{Davis, J.}
\indexauthor{Ireland, M.J.}
\indexauthor{Tuthill, P.G.}
\indexauthor{Tango, W.J.}
\indexauthor{Jacob, A.P.}
\indexauthor{North, J.R.}
\indexauthor{ten Brummelaar, T.A.}

\abstract{The University of Sydney has a long history in optical stellar
 interferometry. The first project, in the 1960s, was the Narrabri Stellar
 Intensity Interferometer, which measured the angular diameters of 32 hot
 stars and established the temperature scale for spectral classes O -
 F. That instrument was followed by the Sydney University Stellar
 Interferometer (SUSI), which is now undergoing a third-generation
 upgrade, to use the multi-wavelength PAVO beam combiner. SUSI operates at
 visible rather than IR wavelengths and has baselines up to 160 m, so it
 is well suited to the study of hot stars. A number of studies have been
 carried out, and more are planned when commissioning of the PAVO system
 is complete. Conversion of the system to allow remote operation will
 allow larger scientific projects to be undertaken. 
}

\resumen{ }

\addkeyword{Stars: angular diameters}
\addkeyword{Stars: fundamental parameters}
\addkeyword{Stars: general}

\begin{document}
\maketitle

\section{the beginning}
\label{sec:NSII}

The concept of intensity interferometry was developed by Hanbury Brown and Twiss in the early 1950s, as a means of obtaining long baselines for high resolution radio astronomy observations, without requiring a phase-stable radio link. It was shown to also work for visible light, in a demonstration using two search-light mirrors (Hanbury Brown \& Twiss 1958), successfully resolving the angular diameter of Sirius.

Building on this proof of concept, the Narrabri Stellar Intensity Interferometer (NSII) was constructed by the University of Sydney and commenced operation in 1963. It observed at a wavelength of 443 nm, and with a maximum baseline of 188 m it was the first instrument able to resolve the angular diameters of hot (main sequence) stars, and the first multi-aperture stellar interferometer since the historic work by Michelson and Pease. The NSII measured the angular diameters of 32 stars with spectral types from O to F (Hanbury Brown et al. 1974), and the temperature scale derived from these observations remains in use today. The NSII also made pioneering measurements of the limb darkening of Sirius, and observations which allowed a full solution for the parameters of the binary system Spica.

Intensity interferometry has the outstanding advantage of immunity from atmospheric turbulence, but on the other hand it is restricted to bright stars. Even with 6.5m diameter mosaic reflectors, the NSII had a practical magnitude limit of $B \simeq 2.5$. It was closed down in 1973 after successfully observing all the stars within its sensitivity limit.

\section{michelson interferometry}
\label{sec:transition}

The possibility of a larger and more sensitive intensity interferometer was investigated in the early 1970s, but by then it was becoming apparent that active tip-tilt correction might deal with the troublesome atmospheric turbulence effects and achieve much greater sensitivity. The group's focus was therefore shifted to the development of a `modern Michelson' stellar interferometer, under the leadership of John Davis.

A prototype interferometer was built in Sydney, and completed in 1985. It
had a fixed NS baseline of 11.4 m, and used 15 cm diameter siderostats. This instrument successfully demonstrated the use of active tip-tilt correction, and measured the angular diameter of Sirius at $\lambda$ 442 nm (Davis and Tango 1986).

\section{the Sydney University Stellar Interferometer (SUSI)}
\label{sec:SUSI}
Funding was granted in 1987 for the construction of a full-scale instrument, known as the Sydney University Stellar Interferometer (SUSI). It is located near Narrabri, on the same site as the CSIRO Australia Telescope radio array. SUSI is a North-South array, with baselines ranging from 5 m to 160 m, although the installed siderostat stations allow future extension to a maximum of 640 m. As a pioneer of the modern long baseline stellar interferometers, SUSI reflects the technology available at the time of design. In particular, it observes in the visible rather than the IR (due to available detectors), uses 20 cm siderostats feeding a 14 cm beam diameter (approximately the Fried length $r_0$ at visible wavelengths, {\it i.e} the maximum beam size suitable for tip-tilt correction) and uses a single baseline at a time. While these design features differ significantly from more recently constructed instruments such as the VLTI or the CHARA array, they nevertheless result in a competitive system which is particularly well suited to studies of hot stars. 

SUSI incorporates what is now standard equipment for optical long-baseline
interferometers - piezo-actuated tip-tilt mirrors, a delay line with laser
metrology, a longitudinal dispersion corrector and a beam combiner. The
instrument was initially set up with the Blue Table beam combiner, using
photomultiplier tube (PMT) detectors ($\Delta \lambda \ 1 - 4$ nm, selected
within $\lambda \lambda \ 430 - 520$ nm) and quad-cell PMT  tip-tilt
detectors. As detector technology advanced, this was superseded by the more
sensitive Red Table combiner ($\Delta \lambda \ 80$ nm, $\lambda \ 700$ nm)
which used Avalanche Photo Diode detectors, a piezo-actuated fringe scanning
system, and a CCD for tip-tilt observations (Davis et al. 2007; see also
Davis 2006). Currently, commissioning is underway for the PAVO beam
combiner, which uses a modern low-light-level CCD detector. With this fast low-noise 2-dimensional detector, it is now possible for PAVO to simultaneously record the band $\lambda \ 520 - 780$ nm in $\simeq 20$ spectral channels. PAVO also uses optimised spatial filtering and pupil segmentation, to reduce the effects of instrumental aberrations and remaining atmospheric phase disturbances. A magnitude limit of $R \simeq 6.5$ is expected. Concurrently we are also implementing full remote control of the instrument. This will allow a much greater fraction of nights to be used, and hence enable substantially larger scientific programs to be undertaken.

The attributes of SUSI which make it particularly suitable for hot star observations are: (a) long baselines, which are needed to resolve the small angular diameters; (b) visible rather than IR wavelengths: this also contributes to resolving small diameter stars, and the stellar flux is stronger than in the IR, while the stellar discs are observed without contributions from dust such as in colliding wind binaries or in circumstellar material; (c) baselines down to 5 m enable stellar systems to be checked for unknown companions, circumstellar emission etc, so visibility curves can be measured without having to {\it assume} that the visibility tends to unity at zero baseline.
%
 
We will not attempt to provide a complete summary of the observational programs that have been carried out with SUSI. Some examples are: $\beta$ Cep binary $\beta$ Cen (Davis et al. 2005); Wolf-Rayet + O star binary $\gamma^2$ Vel (North et al. 2007); $\beta$ Cep/B binary $\sigma$ Sco (North et al. 2007b); triple system including a $\beta$ Cep primary $\lambda$ Sco (Tango et al. 2006); mass determination for single star $\beta$ Hyi (North et al. 2007c);
Be binary $\delta$ Sco (Tango et al. 2009); Cepheid $\ell$ Car (Davis et
al. 2009). 
Further details can be found at http://www.physics.usyd.edu.au/sifa/Main/SUSI. Other work in the group includes masked aperture interferometry (Tuthill et al. 2008).


\begin{thebibliography}
\bibitem[Davis 2006]{2006PASA...23...94} Davis, J.\ 2006, PASA, 23, 94
\bibitem[Davis \& Tango (1986)]{1986Nature...323...234} Davis, J. \& Tango,
W. J.\ 1986, Nature, 323, 234
\bibitem[Davis et al. 2005]{2005MNRAS...356...1362} Davis, J. et al.\ 2005,
MNRAS, 356, 1362
\bibitem[Davis et al. 2007]{2007PASA...24...138} Davis, J. et al.\ 2007,
PASA, 24, 138
\bibitem[Davis et al. 2009]{2009MNRAS...394...1620} Davis, J. et al.\ 2009,
MNRAS, 394, 1620
\bibitem[Hanbury Brown \& Twiss (1958)]{1958Proc. R. Soc. A...248...222}
  Hanbury Brown, R. \& Twiss, R. Q.\ 1958, Proc. R. Soc. A, 248, 222
\bibitem[Hanbury Brown et al. 1974]{1974MNRAS...167...121} Hanbury Brown,
  R., Davis, J. and Allen, L. R.\ 1974, MNRAS, 167, 121
\bibitem[North et al. 2007]{2007MNRAS...377...415} North, J. R., Tuthill,
P. G., Tango, W. J. \& Davis, J.\ 2007, MNRAS, 377, 415
\bibitem[North et al. 2007b]{2007MNRAS...380...1276} North, J. R., Davis, J., Tuthill, P. G., Tango, W. J. \& Robertson, J.G.\ 2007b, MNRAS, 380, 1276
\bibitem[North et al. 2007c]{2007MNRAS...380...L80} North, J. R. et al.\ 2007c, MNRAS, 380, L80
\bibitem[Tango et al. 2006]{2006MNRAS...370...884} Tango, W. J. et al.\ 2006,
MNRAS, 370, 884
\bibitem[Tango et al. 2009]{2009MNRAS...in press} Tango, W. J. et al.\ 2009,
MNRAS in press (arXiv:0811.4004)
\bibitem[Tuthill et al. 2008]{2008ApJ...675...698} Tuthill, P. G. et al.\
2008, ApJ, 675, 698
%
%
%
%
\end{thebibliography}
\end{document}